\documentclass[12pt,aps,preprint,pra,amsmath,amssymb]{revtex4}

\usepackage{epsfig}

\def\la{\langle}
\def\ra{\rangle}

\def\da{^\dagger}

\newcommand{\Op}[1]{{\boldsymbol{\mathrm{\hat{#1}}}}}

\begin{document}

\title{Forcing a unitary transformation by an external field:
comparing two approaches based on optimal control theory.}

\author{Jos\'e P. Palao $^{(a,b)}$ and Ronnie Kosloff $^{(a)}$}

\affiliation{$^{(a)}$ Department of Physical Chemistry and the Fritz 
Haber Research Center for Molecular Dynamics, Hebrew University, 
Jerusalem 91904, Israel\\
$^{(b)}$ Departamento de F\'{\i}sica Fundamental II, Universidad de 
La Laguna, La Laguna 38204, Spain}



\begin{abstract}
\noindent

A quantum gate is realized by specific unitary transformations
operating on states representing qubits. Considering a quantum system
employed as an element in a quantum computing scheme, the task
is therefore to enforce the pre-specified unitary transformation.
This task is carried out by an external time dependent field. Optimal
control theory has been suggested as a method to compute the
external field which alters the evolution of the system such 
that it performs the desire unitary transformation. This study
compares two recent implementations
of optimal control theory to find the field that induces a quantum gate.
The first approach is based on the equation of motion of the unitary
transformation. The second approach generalizes the state to state
formulation of optimal control theory. This work highlight the formal
relation between the two approaches.
\\
\noindent{PACS number(s): 82.53.Kp 03.67.Lx
33.90.+h 32.80.Qk}
\\

\end{abstract}

\maketitle



\section{Introduction.}

Quantum computation is based on implementing selected unitary
transformations representing algorithms \cite{nielsen00}. In 
many physical implementations the unitary transformation is 
generated using an external driving field. 
This driving field has to perform the quantum gate
between, for example, two qubits, without altering
the other levels which represent additional qubits.
This means that the specific unitary transformation
has to address a set of levels in an environment where
other energy levels are present.
An approach to correct this 
undesired coupling to other levels has been suggested 
for specific cases \cite{tian00} but a general solution 
is not known.

The presence of a large number
of levels coupled to the external driving field
is specially relevant in the
implementation of quantum computing in molecular systems
\cite{zadoyan01,amitay02,tesch01}.
Tesh et al. \cite{tesch01} proposed the use of optimal control
theory (OCT) as a possible remedy. OCT \cite{rice00} is a 
well developed approach that allows to obtain the driving field
which induces a desired transition between preselected initial and final
states. However, the dependence on the particular transitions
(characterized by the initial and final states) 
makes the traditional formulation of OCT inappropriate for
that purpose. For 
example, if the unitary transformation relates the initial states 
$\varphi_{ik}$ with the final states $\varphi_{fk}$ (the index $k$
denotes all the relevant states involved in the transformation),
the traditional OCT approach derives an optimal field $\epsilon_k$
for each pair $\{\varphi_{ik},\varphi_{fk}\}$. 
But the fields $\epsilon_k$ obtained are in general different, 
and then the transition induced by $\epsilon_k$ for the initial state 
$\varphi_{ik'}$ won't necessarily give the right final state  
$\varphi_{fk'}$. To implement
a given unitary transformation a single field
$\epsilon$ that relates all the relevant pairs 
$\{\varphi_{ik},\varphi_{fk}\}$ is needed.

An approach to overcome this problem 
was suggested in \cite{palao01poster,palao02}.
The idea is to generalize OCT to deal directly with the evolution
operator. Recently, a different approach using 
simultaneous optimization of several state to state transitions
has been suggested by Tesch and Vivie-Riedle \cite{tesch02}. 
The purpose of this work is to compare the two approaches
and two point the similarities and differences between them.
This comparison has led to new insight into the use of
optimal control theory for quantum systems.


Unitary transformation optimization can be formulated 
in the following way. We consider a quantum 
system with a Hilbert space of dimension $M$, expanded
by an orthonormal basis of states $\{|k\ra\}$ 
($k=1,...,M$). In the following we will assume that 
the $k$ states correspond to the free Hamiltonian proper states. 
The selected unitary transformation is imposed on the subspace
of the first $N$ energy levels of the system ($N\leq M$).
For example, the $N$ levels could correspond
to the physical implementation of the qubit(s) embedded
in a larger system.
The additional levels ($k=N+1,...,M$)
are not directly involved in the target unitary transformation 
and they are generally considered as ``spurious levels'' coupled
to the field. However, it is not always the case:
an example is the proposal of implementing quantum
computation using rovibronic molecular levels.
In the simplest description two electronic surfaces are considered. Two
rovibronic states of one of the electronic surfaces are chosen
as the implementation of a qubit (in this case $N=2$)
and the unitary transformation is implemented using
field induced transitions between the electronic surfaces.
Out of the relevant subspace ($k=3,...,M$) there are
levels residing on both electronic surfaces, spurious in the sense
that any leakage to them would destroy the desired final
result, but at intermediate times they are used as intermediate
storage which allow to carry out the desired unitary transformation
 \cite{palao02,note1}.

The optimization objective is to implement
a selected unitary transformation in the relevant
subspace at a final time $T$.
The target unitary transformation is represented by an
operator in the system Hilbert space and 
denoted by $\Op{O}$. For $N<M$, the matrix 
representation of $\Op{O}$ in the basis $\{|k\ra\}$ 
has two blocks of dimension 
$N\times N$ and $(M-N)\times(M-N)$. The elements
connecting these blocks are zero. 
This structure means that population
at the target time is not transferred between the two
subspaces. Only the $N\times N$ block is relevant for the 
optimization procedure, and the other block remains arbitrary.

The dynamics of the system is generated by the Hamiltonian
$\Op{H}$,
\begin{equation}\label{eq:Hamiltonian}
\Op{H}(t;\epsilon)=\Op{H}_0-\Op{\mu}\,\epsilon(t)\,,
\end{equation}
where $\Op{H}_0$ is the free Hamiltonian, $\epsilon(t)$ is
the driving field and $\Op{\mu}$ is a system operator describing 
the coupling (transition dipole operator in molecular systems). 
Eq. (\ref{eq:Hamiltonian}) can be generalized
to more than one independent driving field,
for example, controlling separately the two components of
the polarization of an electro-magnetic field \cite{brixner01}.
The dynamics of the system at time $t$ is fully specified 
by the evolution operator $\Op{U}(t,0;\epsilon)$.
An optimal field $\epsilon_{opt}$ induces the
target unitary transformation $\Op{O}$ on the system, 
at time $T$ if
\begin{equation}\label{eq:condition}
\Op{U}(T,0;\epsilon_{opt})\,=\,e^{-i\phi(T)}\,\Op{O}\,.
\end{equation}
Eq. (\ref{eq:condition}) implies a condition only on the 
$N\times N$ block of the matrix representation of $\Op{U}$.
The phase $\phi(T)$ is introduced to point out that 
in some cases the target unitary transformation $\Op{O}$ can
be implemented up to an arbitrary global phase.
The phase $\phi$ can be decomposed into two terms, $\phi_1(T)+\phi_2(T)$.
The term $\phi_1$ originates from the arbitrary choice of the
origin of the energy levels which formally means that a term
proportional to the identity operator can always we added to the
Hamiltonian. The phase $\phi_1$ is given by,
\begin{equation}\label{eq:phase}
\phi_{1}(T)=\frac{\sum_{k=1}^M\,E_k\,T}{M\,\hbar}\,,
\end{equation}
where $E_k$ is the energy of the level $k$.
By $\phi_2$ we denoted other contributions to the global
phase due to the structure of the unitary transformation
and its arbitrariness for the levels $k=N+1,...,M$.


\section{Evolution equation approach.}

The optimization approach proposed in \cite{palao01poster,palao02} is
based on Eq. (\ref{eq:condition}) by defining a complex
parameter $\tau$ as
\begin{equation}\label{eq:tau}
\tau(\Op{O};T;\epsilon)
=\sum_{k=1}^N \la k|\Op{O}\da\Op{U}(T,0;\epsilon)|k\ra\,.
\end{equation}
As $\Op{O}$ is a unitary transformation in the relevant
subspace, $\tau$ is a complex number inside a circle 
of radius $N$. Its modulus is equal
to $N$ only when the unitary transformation 
generated by the field $\Op{U}$ is equal to the target
unitary transformation in the relevant subspace, except 
for a possible global phase. 
The modulus of $\tau$ is a measurement of the
fidelity of the target unitary transformation implementation 
by the field \cite{palao02}.
For $N=M$, the sum in Eq. (\ref{eq:tau}) is 
the trace of the operator product. 

In \cite{palao01poster,palao02} the optimization of the real part 
of $\tau$, or the imaginary part, or a linear combination 
of both was suggested as a method to find
the optimal field. For simplicity we will 
consider the real case. The maximization of 
${\rm Re}[\tau]$ can be formulated as a functional optimization.
In this work we use the following form
\begin{eqnarray}\label{eq:functional1}
\bar{J}(\Op{U},\Op{B},\Delta\epsilon)&=&
{\rm Re}\left[\sum_{k=1}^{N} \la k|\,\Op{O}\da 
\Op{U}(T,0;\epsilon_0+\Delta\epsilon)\,|k\ra\right] 
-\lambda \int_{0}^{T} |\Delta\epsilon|^2 dt\nonumber\\
&-&{\rm Re}\left[\,\sum_{k=1}^{N}\int_{0}^{T}
\la k|\,\Op{B}\left(\frac{\partial}{\partial t}+
\frac{i}{\hbar}\Op{H}(t;\epsilon_0+\Delta\epsilon)\,\right)
\Op{U}\,|k\ra\, dt\right]  
\,,
\end{eqnarray}
where $\Op{U}$, $\Op{B}$, $\epsilon_0$, and
$\Delta\epsilon$ depends on $t$. The first term in the 
right-hand-side is the original objective.
In our formulation, the other two terms are 
constrains depending on a reference field $\epsilon_0$
and a field correction $\Delta\epsilon$. The term 
including $|\Delta\epsilon|^2$ minimizes the
total energy of the correction.
The last term introduces the 
dynamics of the system under the field 
$\epsilon_0+\Delta\epsilon$. $\Op{B}$ is an operator 
Lagrange multiplier and $\lambda$ a scalar Lagrange 
multiplier \cite{palao02}. The common 
OCT form for the functional \cite{rice00} 
is recuperated setting $\epsilon_0=0$ and interpreting
$\Delta\epsilon$ as the driving field. However, Eq.
(\ref{eq:functional1}) offers some advantages
in the interpretation of the equations derived from
the functional.
More elaborated constrains are possible \cite{sundermann99,hornung01},
but they are not relevant for the following discussion.

Applying the calculus of variations, $\delta \bar{J}=0$, 
with respect to $\Op{B}$, the Schr\"odinger equation for the
evolution operator of the system is obtained,
\begin{equation}\label{eq:evolutionu}
\frac{\partial\Op{U}}{\partial t} = 
-\frac{i}{\hbar}\Op{H}(t;\epsilon_0+\Delta\epsilon)\,\Op{U}\,,
\end{equation}
with the condition 
$\Op{U}(0,0;\epsilon_0+\Delta\epsilon)=\openone$. The
variation of $\Op{U}$ give the Schr\"odinger evolution
equation for the operator $\Op{B}$,
\begin{equation}\label{eq:evolutionb}
\frac{\partial\Op{B}\da}{\partial t} = 
-\frac{i}{\hbar}\,\Op{H}(t;\epsilon_0+\Delta\epsilon)\,\Op{B}\da\,,
\end{equation}
with the condition 
$\Op{B}\da(T,T;\epsilon_0+\Delta\epsilon)=\Op{O}$. 
$\Op{B}\da$ can be interpreted as the backwards propagation
in time of the target unitary transformation $\Op{O}$. It
is related to $\Op{U}$ by
\begin{equation}\label{eq:uandb}
\Op{B}\da(t,T;\epsilon)=
\Op{U}(t,T;\epsilon)\,\Op{O}\,.
\end{equation}
Eq. (\ref{eq:evolutionu}) and (\ref{eq:evolutionb})
represent the propagation forward and backwards in time 
of the boundary conditions of the problems, that is,
the identity, $\openone$, at time $t=0$ and
the target unitary transformation, $\Op{O}$, at time 
$t=T$. The variation of $\Delta\epsilon$ leads to  an equation
for the correction to the field,
\begin{equation}\label{eq:opfield1}
\Delta{\epsilon(t)} = -\frac{1}{2\,\lambda\,\hbar}\,\,{\rm Im}[\,
\sum_{k=1}^{N} \la k|\,
\Op{B}(t,T;\epsilon_0+\Delta\epsilon)\,\Op{\mu}\,
\Op{U}(t,0;\epsilon_0+\Delta\epsilon)\,
|k\ra]\,.
\end{equation}

Eq. (\ref{eq:opfield1}), more than Eq. (\ref{eq:functional1}),
is the central result of the optimal control procedure
and constitutes the starting point of the iterative
algorithms devoted to determine the optimal field
\cite{palao02}.
When an optimal field $\epsilon_{opt}$ is found $\Delta\epsilon=0$,
and from Eq. (\ref{eq:uandb}) and Eq. (\ref{eq:opfield1}),
\begin{equation}\label{eq:condopt1}
{\rm Im}[\,
\sum_{k=1}^{N} \la k|\,
\Op{O}\da\,\Op{U}\da(t,T;\epsilon_{opt})\,\Op{\mu}\,
\Op{U}(t,0;\epsilon_{opt})\,
|k\ra]\,=0\,.
\end{equation}

Eq. (\ref{eq:condopt1}) constitutes a condition for the optimal
fields for the first approach.
It is the base for the following analysis.
Let us denoted by $\tilde{\epsilon}$ a field that generated
the target unitary transformation up to a global phase,
$\Op{U}(T,0;\tilde{\epsilon})=e^{-i\phi}\,\Op{O}$.
Using 
\begin{equation}\label{eq:uandu}
\Op{U}(t,0,\epsilon)=\Op{U}(t,T,\epsilon)\Op{U}(T,0,\epsilon)\,,
\end{equation}
the diagonal block structure of the matrix 
representation of $\Op{O}$ in the basis $\{|k\ra\}$,
and the relation $\la k|\Op{\mu}|k\ra=0$, it is
found that the left-hand-side of Eq. (\ref{eq:condopt1})
gives
\begin{equation}\label{eq:opgood1}
{\rm Im}[\,
\sum_{k=1}^{N} e^{-i\phi} \la k|\,
\Op{U}(t,T;\tilde{\epsilon})\,\Op{O}\,\Op{O}\da\,
\Op{U}\da(t,T;\tilde{\epsilon})\,\Op{\mu}\,
|k\ra]\,=0\,.
\end{equation}
This result implies that any field inducing the target
unitary transformation up to a global phase
is a possible optimal solution of the
optimization algorithm based on Eq. (\ref{eq:opfield1}).
The convergence to such a solution will depend on the
particular numerical implementation. It must be remarked
that Eq. (\ref{eq:condopt1}) is a necessary but not
a sufficient condition for the optimal field.
For example, let us consider a target unitary transformation
diagonal in the basis $\{|k\ra\}$ denoted by $\Op{O}_D$.
The unitary transformation generated
by the free Hamiltonian  $\Op{U}(t_1,t_2;\epsilon=0)$ is 
also diagonal in that basis, and then,
\begin{equation}\label{eq:opbad1}
{\rm Im}[\,
\sum_{k=1}^{N} \la k|\,
\Op{O}_D\da\,\Op{U}\da(t,T;\epsilon=0)\,\Op{\mu}\,
\Op{U}(t,0;\epsilon=0)\,
|k\ra]\,=0\,.
\end{equation}
However, in general $\Op{U}(T,0;\epsilon=0)$ is not equal to
the target $\Op{O}_D$. These spurious solutions to the
optimization can be avoided with a different choice of the
initial guess for the algorithm.


In optimization procedures based on Eq. (\ref{eq:opfield1}) 
the full operator propagation in Eq. 
(\ref{eq:evolutionu}) and (\ref{eq:evolutionb})
is not needed since only the action of $\Op{U}$ and 
$\Op{O}$ on the states $|k\ra$ in the relevant
subspace ($k=1,...,N$) appears.
Then only the first $N$ rows of the matrix operator 
representations are propagated.
Denoting by $U^k$ ($(B\da)^k$) the $k$ row of the matrix representation
of $\Op{U}$ ($\Op{B}\da$) in the basis $\{|k\ra\}$, and being $U^k_j$
($(B\da)^k_j$) the $j$ element of the row ($j=1,...,M$), the evolution 
equations (\ref{eq:evolutionu}) and (\ref{eq:evolutionb})
take the form,
\begin{eqnarray}\label{eq:evolutionrow}
\frac{\partial U^k(t)}{\partial t} &=& 
-\frac{i}{\hbar}H(t)\,U^k(t)\,,\nonumber\\
\frac{\partial (B\da)^k(t)}{\partial t} &=& 
-\frac{i}{\hbar}H(t)\,(B\da)^k(t)\,,
\end{eqnarray}
with the conditions $U^k_j(t=0)=\delta_{jk}$
and $(B\da)^k_j(t=T)=O^k_j$ respectively, being
$O^k_j$ the matrix elements of the target unitary
transformation $\Op{O}$ and $H(t)$ the matrix
representation of $\Op{H}$.
The $2N$ evolution equations ($k=1,...,N$) are equivalent to 
the propagation of $2N$ states of the
system. The advantage is that in this case only $2(N\times M)$ 
elements are propagated instead of the $2(M\times M)$ in Eq.
(\ref{eq:evolutionu}) and Eq. (\ref{eq:evolutionb}).



\section{State to state approach.}

The state to state approach
\cite{tesch02} is based on the simultaneous optimization of 
$N$ transitions between pairs of initial and final 
states. These pairs of states $\{\varphi_{il},\varphi_{fl}\}$, ($l=1,...,N$) 
are related by the target unitary transformation, 
$|\varphi_{fl}\ra=\Op{O}\,|\varphi_{il}\ra$.
The objective is formulated as
\begin{equation}\label{eq:eta}
\eta(\Op{O};T;\epsilon)
=\sum_{l=1}^{N} |\la \psi_{il}(T;\epsilon)|\varphi_{fl}\ra|^2\,,
\end{equation}
where $\psi_{il}(T;\epsilon)$ is the state at the target time
that evolves with the driving field $\epsilon(t)$ and the initial 
condition $\psi_{il}(t=0)=\varphi_{il}$. 
$\eta$ is a positive real number and its  maximum value $\eta=N$ 
is reached when all the initial states $\varphi_{il}$
are driven by the field to the correct final states $\varphi_{fl}$. 
The task of obtaining the optimal field is equivalent to the maximization
of $\eta$.
The set of initial states $\varphi_{il}$ must be chosen
carefully. In order to account for all the possible transitions
the states $\varphi_{il}$ have to represent the relevant subspace. 
However, the choice of an orthonormal basis 
could produce undesired results.
Let us denoted by $\{\tilde{\varphi}_{il}\}$ an orthonormal basis
of the relevant subspace
and by $\Op{D}$ an arbitrary unitary transformation  
diagonal in that basis. The product $\Op{O}\,\Op{D}$ 
is also a unitary transformation. If 
$\epsilon_O$ and $\epsilon_{OD}$ are fields that
generate $\Op{O}$ and $\Op{O}\,\Op{D}$ at time $T$ respectively,
it is found that the same optimization objective is obtained,
\begin{equation}
\eta_\perp(\Op{O};T;\epsilon_O)=\eta_\perp(\Op{O};T;\epsilon_{OD})\,,
\end{equation}
where $\perp$ denotes that $\eta$ was evaluated using an orthonormal 
basis.
Then any algorithm based on
$\eta$ and using an orthonormal basis would find 
the optimal field corresponding to any of the 
possible targets $\Op{O}\,\Op{D}$ 
($\Op{O}$ is a particular case when $\Op{D}$
is the identity operator).
The reason is that $\eta$ is
sensitive to the modulus of the projection of
each pair $\{\varphi_{il},\varphi_{fl}\}$ but not to the
relative phases between them.
The phase problem can be overcome with a careful choice 
of the states $\varphi_{il}$. 
Keeping the first $N-1$ states of the basis and
substituting the $N$ state by 
$\sum_{l=1}^{N}\tilde{\varphi}_{il}/\sqrt{N}$,
the maximum condition is achieved only when the
field induces the target unitary transformation up to
a possible global phase.


The optimization is formulated as the maximization
of the functional
\begin{eqnarray}\label{eq:functional2}
\bar{K}(\psi_{ik},\psi_{fk},\Delta\epsilon)&=&
\sum_{l=1}^{N} |\la\psi_{il}(T;\epsilon_0
+\Delta\epsilon)|\varphi_{fl}\ra|^2
-\lambda \int_{0}^{T} |\Delta\epsilon(t)|^2 dt
\nonumber\\
&-&2\,{\rm Re}\left[\sum_{l=1}^{N}
\la\psi_{il}(T;\epsilon_0+\Delta\epsilon)|\varphi_{fl}\ra
\int_{0}^{T} \la\psi_{fl}|
\left(\frac{\partial}{\partial t}
+\frac{i}{\hbar}\Op{H}(t;\epsilon_0
+\Delta\epsilon)\,\right)
|\psi_{il}\ra\,dt\right]
\,,\nonumber\\
&&
\end{eqnarray}
where $\psi_{il}$, $\psi_{fl}$, $\epsilon_0$ and $\Delta\epsilon$
depend on time. The form in Ref. \cite{tesch02} is recuperated
setting $\epsilon_0=0$ and interpreting $\Delta\epsilon$ as the
driving field.

The variations with respect to $\psi_{li}$ and
$\psi_{lf}$ ($l=1,...,N$) lead to the following $2N$ equations
\begin{equation}\label{eq:evolutionpi}
\frac{\partial}{\partial t}|\psi_{il}\ra= 
-\frac{i}{\hbar}\Op{H}(t;\epsilon_0+\Delta\epsilon)\,|\psi_{il}\ra\,,
\end{equation}
with the condition $|\psi_{il}(t=0)\ra=|\varphi_{il}\ra$, and

\begin{equation}\label{eq:evolutionpf}
\frac{\partial}{\partial t}|\psi_{fl}\ra= 
-\frac{i}{\hbar}H(t;\epsilon_0+\Delta\epsilon)\,|\psi_{fl}\ra\,,
\end{equation}
with the condition $|\psi_{fk}(t=T)\ra=|\varphi_{fk}\ra$.
The variation respect to $\Delta\epsilon$ leads to the equation
for the correction to the field,
\begin{equation}\label{eq:opfield2}
\Delta{\epsilon_{ss}(t)} = -\frac{1}{\lambda\,\hbar}\,\,{\rm Im}\left[\,
\sum_{l=1}^{N}\la\psi_{fl}(T;\epsilon_0+\Delta\epsilon)|\varphi_{fl}\ra
\la\psi_{fl}(t;\epsilon_0+\Delta\epsilon)|
\,\Op{\mu}\,|
\psi_{il}(t;\epsilon_0+\Delta\epsilon)\ra
\right]\,.
\end{equation}
%

Eq. (\ref{eq:evolutionpi}) and (\ref{eq:evolutionpf}) 
are formally equivalent to Eq. (\ref{eq:evolutionrow}),
propagating forward or backward on time the boundary 
conditions of the problem.
Then the main difference between the two approaches is found in
the expression for the correction to the field,
Eq. (\ref{eq:opfield1}) and Eq. (\ref{eq:opfield2}).
The connection between the two expressions is established
by rewriting the states $\varphi$ and $\psi$ as
\begin{eqnarray}\label{eq:relation}
|\varphi_{il}\ra &\rightarrow& |l\ra, \nonumber\\
|\varphi_{fl}\ra &\rightarrow& \Op{O}\,|l\ra, \nonumber\\
|\psi_{il}(t;\epsilon)\ra &\rightarrow& 
\Op{U}(t,0;\epsilon)\,|l\ra, \nonumber\\
|\psi_{fl}(t;\epsilon)\ra &\rightarrow& \Op{B}\da(t,T;\epsilon)\,|l\ra\,
\end{eqnarray}
and Eq. (\ref{eq:opfield2}) as,
\begin{equation}
\Delta\epsilon_{ss}(t) = -\frac{1}{\lambda\,\hbar}\,\,{\rm Im}\left[\,
\sum_{l=1}^{N}\la l|\Op{U}\da(T,0;\epsilon_0+\Delta\epsilon)\,\Op{O}|l\ra\,
\la l|\Op{B}(t,T;\epsilon_0+\Delta\epsilon)\,\Op{\mu}\,
\Op{U}(t,0;\epsilon_0+\Delta\epsilon)|l\ra
\right]\,.
\end{equation}
This expression is formally equivalent to 
Eq. (\ref{eq:opfield1}) except for the factor
$\la l|\Op{U}\da(T,0,\epsilon_0+\Delta\epsilon)\,\Op{O}|l\ra$.
(The different factor $1/2$ can be removed with a redefinition of
$\lambda$). 
Omitting this factor the two approaches would be completely
equivalent: in that case the careful choice of the 
initial state set wouldn't be necessary and the two approaches
would lead to the same set of equations.
In general the equivalence is only formal due to the 
additional factor and the different set
of states $\{|k\ra\}$ and $\{|l\ra\}$. In the following
we will chose $|l\ra=|k\ra$ for $l=1,...,N-1$ and
$|l=N\ra=\sum_{k=1}^N|k\ra/\sqrt{N}$. The condition for 
the optimal field ($\Delta\epsilon=0$) is obtained
using Eq. (\ref{eq:opfield2}) and (\ref{eq:uandb}),
\begin{equation}\label{eq:condopt2}
{\rm Im}\left[\,
\sum_{l=1}^{N}\la l|\Op{U}\da(T,0;\epsilon_{opt})\,\Op{O}|l\ra\,
\la l|\Op{O}\da\,\Op{U}\da(t,T;\epsilon_{opt})\,\Op{\mu}\,
\Op{U}(t,0;\epsilon_{opt})|l\ra
\right]\,=\,0\,.
\end{equation}
Eq. (\ref{eq:condopt2}) is a 
necessary but not sufficient condition for the optimal
field, as is also Eq. (\ref{eq:condopt1}).
Denoting as before by $\tilde{\epsilon}$ a driving field 
generating the target unitary transformation up to a global
phase, the left-hand-side of (\ref{eq:condopt2}) gives,
\begin{equation}\label{eq:opgood2}
{\rm Im}\left[\,
\sum_{l=1}^{N}\,
\la l|\Op{O}\da\,\Op{U}\da(t,T;\tilde{\epsilon})\,\Op{\mu}\,
\Op{U}(t,T;\tilde{\epsilon})\,\Op{O}|l\ra
\right]\,=\,0\,.
\end{equation}
(The property ${\rm Im}[\la\Psi|\Op{\mu}|\Psi\ra]=0$ for Hermitian
operator $\Op{\mu}$ was used). Contrary to
Eq. (\ref{eq:opgood1}), the phase  
$\phi$ doesn't appears in Eq. (\ref{eq:opgood2}) due to 
the factor
$\la l|\Op{U}\da(T,0,\tilde{\epsilon})\,\Op{O}|l\ra$ 
correcting each term in the sum.
Using the same arguments leading to Eq. (\ref{eq:opbad1}),
it can be shown that the left-hand-side of Eq. (\ref{eq:condopt2})
is null for $\epsilon=0$ when the target unitary transformation
is diagonal.
The previous results imply that any field that generates the
target unitary transformation up to a global phase fulfills the 
condition in Eq. (\ref{eq:condopt2}) and in addition spurious
solutions could be found, as in the first approach.

\section{Conclusions}

In this study we have employed a modified formulation of 
OCT which uses the correction to the driving field $\Delta\epsilon$
as central element. $\Delta\epsilon=0$ is a necessary but
not sufficient condition for obtaining the objective. 
The relations derived from this condition allow a better 
analysis of the control equations, in particular, the similarities 
and differences between the two approaches. 
Optimal solutions of the evolution equation approach will be also
optimal solutions of the state to state approach.
In this sense both approaches are equivalents.
The difference between them is a term in the state
to state approach which modified the phases.
This term can cause a phase ambiguity
in the target unitary transformation. A careful 
choice of the initial set of states can solve this
problem.



\section{Acknowledgments}

We want to thank Lajos Diosi and Zohar Amitay for
many useful discussions.
This work was supported by the Spanish 
MCT BFM2001-3349 and the Israel Science Foundation. 
The Fritz Haber Center is supported by the Minerva 
Gesellschaft f\"ur die Forschung, GmbH M\"unchen, Germany.




\begin{thebibliography}{10}

\bibitem{nielsen00} M. A. Nielsen and I. L. Chuang, 
{\it ``Quantum computation and quantum information''},
(Cambridge University Press, 2000).

\bibitem{tian00}L. Tian and S. Lloyd, Phys. Rev. A
{\bf 62}, 050301 (2000).

\bibitem{zadoyan01} R. Zadoyan, D. Kohen, D. A. Lidar, and
V. A. Apkarian, Chem. Phys. {\bf 266}, 323 (2001).

\bibitem{amitay02} Z. Amitay, R. Kosloff, and S. R. Leone,
Chem. Phys. Lett. {\bf 359}, 8 (2002);
J. Vala, Z. Amitay, B. Zhang, S. R. Leone, and
R. Kosloff, quant-ph/0107058.

\bibitem{tesch01} C. M. Tesch, L. Kurtz, and R. de Vivie-Riedle,
Chem. Phys. Lett. {\bf 343}, 633 (2001).

\bibitem{rice00} S. A. Rice and M. Zhao, 
{\it ``Optimal Control of Molecular Dynamics''},
(John Wiley \& Sons, Inc., New York, 2000), and references
therein.

\bibitem{palao01poster} J. P. Palao and R. Kosloff, 
in proceedings of the conference 
{\it ``Coherent Control and Cold Molecules''}, Gif-sur-Yvette, 
France, October 21-25, 2001.
 
\bibitem{palao02} J. P. Palao and R. Kosloff, 
quant-ph/0204101.

\bibitem{tesch02} C. M. Tesh and R. de Vivie-Riedle,
quant-ph/0208025.

\bibitem{note1} The use of intermediate states
introduces two different times
scales in the problem, one related with the direct 
transition between electronic surfaces and the other 
related with the indirect transition between the
qubit levels. It is found that the 
optimization algorithms convergence is slower in this case. 
The advantage is that there are
transitions between electronic surfaces in the visible
region, for which the shaping pulse technology is well
developed, making feasible the experimental implementation
of the optimized field.

\bibitem{brixner01} T. Brixner and G. Gerber, Op. Lett. {\bf 26},
557 (2001).

\bibitem{sundermann99} K. Sundermann and R. de Vivie-Riedle,
J. Chem. Phys. {\bf 110}, 1896 (1999).

\bibitem{hornung01} T. Hornung, M. Motzkus, and R. de Vivie-Riedle,
J. Chem. Phys. {\bf 115}, 3105 (2001); Phys. Rev. A {\bf 65},
032514 (2002).







 
\end{thebibliography}
\end{document}